\documentclass[journal,draftclsnofoot,onecolumn,12pt,twoside]{IEEEtran}
\usepackage{times,amsmath,epsfig}
\usepackage{cite}
\usepackage{graphicx}
\usepackage{psfrag}
\usepackage{subfigure}
\usepackage{url}
\usepackage{float}
\usepackage{stfloats}
\usepackage{amsmath}
\usepackage{amssymb}

\begin{document}

\title{Full-duplex Amplify-and-Forward Relaying: \\Power and Location Optimization }

\author{\IEEEauthorblockN{Shuai Li\IEEEauthorrefmark{1}, Kun Yang\IEEEauthorrefmark{1}, Mingxin Zhou\IEEEauthorrefmark{1}, Jianjun Wu\IEEEauthorrefmark{1}, Lingyang Song\IEEEauthorrefmark{1},\\ Yonghui Li\IEEEauthorrefmark{2}, and Hongbin Li\IEEEauthorrefmark{1}} \IEEEauthorblockA{\\\IEEEauthorrefmark{1} School of Electronics Engineering and Computer Science,\\ Peking University, Beijing, China\\ \IEEEauthorrefmark{2}School of Electrical and Information Engineering,\\ The University of Sydney, Australia\\  }

\thanks{Copyright (c) 2015 IEEE. Personal use of this material is permitted. However, permission to use this material for any other purposes must be obtained from the IEEE by sending a request to pubs-permissions@ieee.org. }
}

\maketitle
 \thispagestyle{empty}
\pagestyle{empty}

\vspace{15mm}
\begin{abstract}
In this paper, we consider a full-duplex (FD) amplify-and-forward (AF) relay system and optimize its power allocation and relay location to minimize the system symbol error rate (SER).
We first derive the asymptotic expressions of the outage probability and SER performance by taking into account the residual self interference (RSI) in FD systems.
We then formulate the optimization problem based on the minimal SER criterion.
Analytical and numerical results show that optimized relay location and power allocation can greatly improve system SER performance, and the performance floor caused by the RSI can be significantly reduced via optimizing relay location or power allocation.

\end{abstract}


\section{Introduction}
Full-duplex (FD) has emerged as a promising technology to increase the spectrum efficiency of next-generation wireless networks~\cite{MJain,DB,ASabharwal}.
Recently, FD were deployed in the relay networks to improve the performance of the relay networks~\cite{Ju}.
In the FD relay networks, the FD relays receive and transmit the information over the same frequency at the same time, rather than over two orthogonal channels as in half-duplex~(HD) relay systems~\cite{GL}.
The performance of such a system is seriously affected by the self interference caused by the signal leakage from the transmit antennas to the receive antennas at the relay node~\cite{RiihonenWCNC}.
An optimal relay selection scheme with dynamic switching between FD and HD has been proposed to mitigate the influence of self interference~\cite{Ik}.
To enhance the performance of FD relay system, various approaches, such as self interference mitigation~\cite{RiihonenTSP,MD,EE}, antenna selection~\cite{Antenna}, transceiver beamforming~\cite{JHL}, have been recently developed.

In wireless relay networks, resource allocation is an effective approach to improving the spectral efficiency and error performance~\cite{MDo}.
The two-dimensional resource allocation optimization, including energy optimization and location optimization, was studied to provide system symbol error rate (SER) advantages and achieve the minimum SER for the HD relay networks in~\cite{wcho}.
Optimal relay location and power allocation for HD relay networks has been studied intensively in previous literature ~\cite{MOH,CY,RC}.
However, the previous optimization schemes for HD relay networks can not be directly applied to FD relay networks due to the impact of residual self interference (RSI)~\cite{BoYu2}.
The distribution of the signal to interference plus noise ratio (SINR) and corresponding power/location optimization for the FD relay system are different from conventional HD communication, thus new optimization schemes are needed.

In practice, optimal power allocation is an effective way to improve the system performance.
In~\cite{RiihonenTWC}, the combination of power adaptation and mode selection for maximizing average and instantaneous spectral efficiency based on the SINR was studied for a two-hop relay system.
The capacity and respective optimal power allocation for FD dual-hop system under the RSI modeled as Gaussian distribution was analyzed in~\cite{Rodriguez}.
Two optimal power allocation strategies which are designed based on statistical channel state information (CSI) and instantaneous CSI for FD relay networks with amplify-and-forward protocol and total transmit power constraint, were proposed to minimize the outage probability in~\cite{TPDo}.
An adaptive power allocation scheme combined with relay and mode selection was introduced in the AF multiple relay network to mitigate the RSI and reduce the performance floor in the high SNR region~\cite{Yang}.

The optimization of relay location is also essential to enhance the system performance.
The optimal relay location for conventional HD relay systems has been studied extensively~\cite{HVZhao}.
The optimal relay location for the FD relay systems with decode-and-forward (DF) relaying protocol was investigated in~\cite{BoYu} based on the minimal outage probability criterion.
However, no closed-formed solution for the optimal relay location with DF protocol has been derived due to the complex SINR distribution of the received signal,
and a suboptimal solution was provided for the low RSI case in~\cite{BoYu2}.

In this paper, we study the joint power allocation and relay location optimization problem for FD relay systems where the source, relay and destination are in a straight line. 
We prove that the SER performance is convex in terms of power allocation and relay location variables.
We assume that the RSI is considered as Rayleigh distribution in our analysis \cite{MD,Tk}.
Due to the RSI at the relay, the SINR expression at the destination in the FD relay systems becomes very complicated and essentially different from that in the HD relay systems.
To tackle this issue, we need to derive a new probability distribution of SINR at the destination, which is a non-trivial task when taking the variation of RSI into the consideration.
By using some high SNR approximations, we derive a closed-form CDF expression of the end-to-end SINR, and obtain the asymptotic expressions of the outage probability and average SER.
The derived expression of outage probability is different from the existing literatures in \cite{Ik, Antenna} due to the different analytical method.
Based on the derived asymptotic expressions, new expressions for the optimal location and power allocation solutions are derived, respectively.
For the joint optimization problem, we propose a suboptimal solution in high transmit power region. 
The analytical results are validated by numerical simulations.

The rest of the paper is organized as follows.
In Section II, we present the system model.
In Section III, the asymptotic expressions for the outage probability and SER performance of FD relay system are provided.
The joint optimization, power allocation and location optimization problems based on the minimal SER criterion are formulated and analyzed in Section IV.
Numerical results are provided in Section V.
Main conclusions are drawn in Section VI.

\IEEEpeerreviewmaketitle
 \thispagestyle{empty}
\section{System Model}

In this paper, we consider a two-hop relay systems with one source (S), one destination (D), and one AF relay node (R).
We consider the relay deployment scenario of dead spot where the direct link between the source and destination is blocked by physical obstacles or barriers and the relay is deployed to achieve coverage extension for users in coverage holes \cite{R1,R2,R3}.
We assume the direct link between the source and the destination is strongly attenuated and communication can be only established via the relay.
The relay is equipped with two antennas, one for reception and the other for transmission.
The relay works in the FD transmission mode, and both source and relay use the same time-frequency resource.
The distance between the source-relay, relay-destination, source-destination are denoted as $D_{SR}$, $D_{RD}$, $D_{SD}$.
The sum distance between the source-relay and relay destination is denoted as $D =D_{SR}+D_{RD}$.
As the direct link is strongly attenuated, the relay can not be placed in the line of source-destination and thus $D > D_{SD}$.
The channels for the source-relay, relay-destination, source-destination are denoted as $h_{SR}, h_{RD}, h_{SD}$ respectively, and $h_{SD} \approx 0$.

\begin{figure}[!t]
\centering
\includegraphics[width=5in]{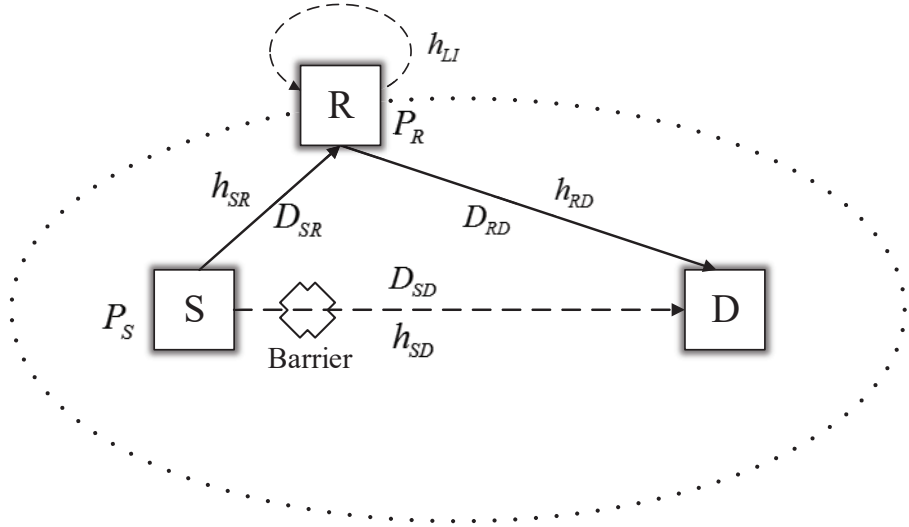}
\caption{\label{fig:model}System model of AF FD relay}
\end{figure}

The received signal $y_R$ at the relay is given as
\begin{equation}\label{eq:relay_signal}
y_{R} = h_{SR} \sqrt{P_S} x_{S} + h_{LI} \sqrt{P_R} x_{R} + n_{R} ,
\end{equation}
where $h_{LI}$ is the RSI at the FD relay.
$x_S$ and $x_R$ are the signal with the unit power transmitted from the source and the relay respectively.
$P_S$ and $P_R$ is the transmit power of the source and relay, and ${P_S} + {P_R} = P$.
$n_R$ is the additive white Gaussian noise (AWGN) at the relay with the variance $N_0$.

Upon receiving the signal from the source, the relay uses the AF protocol to forward the following signal,
\begin{equation}
x_R  = \beta y_R ,
\end{equation}
where $\beta$ is the power amplification factor to ensure that the average power of signal $x_R$ satisfies the following power constraint,
\begin{equation}\label{eq:power_limit}
\mathbb{E} \big[ |x_R|^2 \big] = \beta^2 \Big( |h_{SR}|^2 P_S + |h_{LI}|^2 P_R + \sigma^2 \Big) \leq 1 .
\end{equation}

The received signal at the destination is given by
\begin{equation}\label{eq:received_signal}
{y_D} = {h_{RD}}\sqrt {{P_R}} {x_R} + {h_{SD}}\sqrt {{P_S}} {x_S} + {n_D},
\end{equation}
where $n_D$ is AWGN with mean zero and variance $N_0$ at the destination.

Therefore, the end-to-end SINR can be expressed as
\begin{align}\label{eq:e2eSINR}
{\gamma _{SINR}} &= \frac{{\frac{{{\gamma _{SR}}}}{{{\gamma _{LI}} + 1}}\frac{{{\gamma _{RD}}}}{{{\gamma _{SD}} + 1}}}}{{\frac{{{\gamma _{SR}}}}{{{\gamma _{LI}} + 1}} + \frac{{{\gamma _{RD}}}}{{{\gamma _{SD}} + 1}} + 1}} \nonumber\\
 &\approx \frac{{{\gamma _{SR}}{\gamma _{RD}}}}{{{\gamma _{SR}} + ({\gamma _{RD}} + 1)({\gamma _{LI}} + 1)}},
\end{align}
where $\gamma_{SR} = P_{S}|h_{SR}|^2/N_0$, $\gamma_{RD} = P_{R}|h_{RD}|^2/N_0$, $\gamma_{SD} = P_{S}|h_{SD}|^2/N_0$, and $\gamma_{LI} = P_{S}|h_{LI}|^2/N_0$.
As we assume the direct link is strongly attenuated and $h_{SD} \approx 0$, $\gamma_{SD}$ is omitted from (5) for approximation.
The source-relay, relay-destination and self interference channel are modelled as independent Rayleigh flat fading.
The SNR of the channel link $h_{SR}$, $h_{RD}$ and $h_{LI}$ are exponentially distributed with mean $\lambda_{SR}$, $\lambda_{RD}$ and $\lambda_{LI}$, respectively.
The average channel SNRs can be expressed as $\lambda_{SR} = P_S D_{SR}^{-v}$, $\lambda_{RD} = P_R D_{RD}^{-v}$, and ${\lambda _{LI}} = \varepsilon {P_R}$, where $v$ denotes the path loss exponent of the wireless channel and $\varepsilon$ denotes the RSI level.

\section{Outage Probability and SER}

In this section, we derive the asymptotic expressions of the outage probability and SER performance for FD relay systems. The relay location ratio and power allocation ratio are defined as ${\rho _D} = {D_{SR}}/D$ and ${\rho _{\lambda  } } = {P_S}/P$, respectively.

\subsection{Outage Probability}

Due to the RSI at the relay, the SNR expressions at the destination in the FD relay systems is essentially different from that in the HD relay systems. In Theorem 1, we derive a closed-form asymptotic CDF expression of end-to-end SINR.

\textbf{\emph{Theorem 1: }} The asymptotic CDF expression of the end-to-end SINR can be calculated as
\begin{equation}
\label{eq:CDF_onerelay}
F(x) =
1-\frac{e^{- (\frac{1}{\lambda_{SR}} + \frac{1}{\lambda_{RD}}) x}}{1+\eta x} \frac{2x}{\sqrt{\lambda_{SR}\lambda_{RD}}} K_1 \left( \frac{2x}{\sqrt{\lambda_{SR}\lambda_{RD}}} \right),
\end{equation}
where $K_1(\cdot)$ is the first order modified Bessel function of the second kind~\cite{Abramowitz}, and $\eta = \lambda_{LI}/\lambda_{SR}$.

\emph{\textbf{Proof: }}
See Appendix A.$\hfill\blacksquare$

\subsection{Average SER}

The average SER can be calculated by
\begin{equation}\label{eq:ser_int}
\overline{SER} = \alpha \mathbb{E}  \left[Q\left(\sqrt{\beta \gamma} \right)\right]= \frac{\alpha\sqrt{\beta}}{2\sqrt{2\pi}} \int\limits_0^\infty { \frac{1}{\sqrt{t}} F(t) e^{-\frac{\beta}{2}t } } dt,
\end{equation}
where $F(\cdot)$ is the CDF of the end-to-end SINR given in ~\eqref{eq:CDF_onerelay}, and $Q(\cdot)$ is the Gaussian \emph{Q}-Function\cite{Abramowitz}.
The parameters $(\alpha, \beta)$ depend on the modulation formats, e.g., $\alpha = 1$, $\beta = 2$ for BPSK modulation.

\textbf{\emph{Theorem 2: }} The asymptotic end-to-end SER expression can be approximated as
\begin{equation}
\label{eq:ser_approx}
SER \approx \frac{1}{2} - \sum\limits_{i = 0}^{{N_I}} {{I_i}} ,
\end{equation}
where
\begin{small}
\begin{equation}\label{eq:I}
I_{i} = C_i
\frac{2\alpha \sqrt{2\beta}}{\lambda_{SR}\lambda_{RD} }
 \frac{A_i \eta^{2i}} {X_i^{2i+\frac{5}{2}}}
{}_2{F_1} \left( 2i+\frac{5}{2}, \frac{3}{2}; 2i+2; \frac{Y_i}{X_i} \right),
\end{equation}
\end{small}
and
\begin{eqnarray}
C_i &=&  \frac{\Gamma(2i+\frac{5}{2})\Gamma(2i+\frac{1}{2})}{(2i+1)!}\nonumber\\
X_i &=& \frac{\beta}{2} + \eta B_i + (\frac{1}{\sqrt{\lambda_{SR}}} + \frac{1}{\sqrt{\lambda_{RD}}} )^2 ,\nonumber\\
Y_i &=& \frac{\beta}{2} + \eta B_i + (\frac{1}{\sqrt{\lambda_{SR}}} - \frac{1}{\sqrt{\lambda_{RD}}} )^2 ,
\end{eqnarray}
where the function ${}_2F_1(\cdot)$ denotes the hypergeometric function~\cite{Abramowitz}, and the parameters $A_1$ and $B_i$ are presented in~\eqref{eq:para_ser}, ${{N_I}}$ denotes the number of $({A_i},{B_i})$ pairs used for approximation.

\emph{\textbf{Proof: }}
See Appendix B.$\hfill\blacksquare$

At high transmit power, \eqref{eq:I} can be reduced as
\begin{equation}\label{eq:Iapprox}
I_i \to \frac{\alpha\sqrt{\beta}}{2\sqrt{2\pi}} {A_i \eta^{2i}} \Gamma(2i+\frac{1}{2}) (\frac{\beta}{2} + \frac{1}{\lambda_{SR}} + \frac{1}{\lambda_{RD}} + {B_i\eta} )^{-\frac{4i+1}{2}}.
\end{equation}

Indicated by~\eqref{eq:Iapprox}, $I_i\propto \eta^{2i}$.
Therefore, the SER performance is mainly determined by item $I_0$
\begin{equation}\label{eq:I0}
  {I_0} = \frac{{\alpha \sqrt \beta  }}{{2\sqrt {2\pi } }}\Gamma (\frac{1}{2}){(\frac{\beta }{2} + \frac{1}{{{\lambda _{SR}}}} + \frac{1}{{{\lambda _{RD}}}}{\rm{ + }}{B_0}\eta )^{ - \frac{1}{2}}},
\end{equation}

At high SNR, the average SER can be approximated as
\begin{equation}
\label{eq:serhighhigh}
SER \approx \frac{1}{2} - \frac{{\alpha \sqrt \beta  }}{{2\sqrt {2\pi } }}\Gamma (\frac{1}{2}){(\frac{\beta }{2} + \frac{1}{{{\lambda _{SR}}}} + \frac{1}{{{\lambda _{RD}}}}{\rm{ + }}{B_0}\eta )^{ - \frac{1}{2}}},
\end{equation}

With \eqref{eq:serhighhigh}, the SER optimization problem can be simplified to calculate the optimal power allocation and relay location which minimize the value of $f({\rho _\lambda } ,{\rho _D} )$
\begin{equation}\label{eq:f}
  f({\rho _\lambda } ,{\rho _D} ) \hspace{-1mm} =\frac{\beta }{2} + \frac{1}{{{\lambda _{SR}}}} + \frac{1}{{{\lambda _{RD}}}}{\rm{ + }}\eta \hspace{-1mm}= \frac{\beta }{2} + \frac{{1 + \varepsilon {P_R}}}{{{P_S}}}D_{SR}^v + \frac{{D_{RD}^v}}{{{P_R}}}.
\end{equation}

In the high SNR region, the SER of the FD relay system encounters the performance floor. When SNR comes to infinity, the performance floor can be expressed as
\begin{equation}\label{eq:performance floor}
SE{R_{SNR -  > \infty }} = \frac{1}{2} - \frac{{\alpha \sqrt \beta  }}{{2\sqrt {2\pi } }}\Gamma (\frac{1}{2}){(\frac{\beta }{2} + \varepsilon \frac{{{P_R}}}{{{P_S}}}D_{SR}^v)^{ - \frac{1}{2}}},
\end{equation}
when RSI level $\varepsilon = 0$, $SER_{{SNR -  > \infty }}=(1-\alpha)/2$, indicating that the performance floor can be removed with perfect self interference cancellation.

With \eqref{eq:performance floor}, we know that the performance floor is determined by RSI level $\varepsilon$, the relay location ratio and the power allocation ratio, which means that optimizing the relay location and power allocation ratio is a feasible way to mitigate the performance floor caused by imperfect interference cancellations.

\section{Power Allocation And Relay Location Optimization}

In this section, we investigate the power allocation and relay location optimization to minimize SER for the FD relay network.

\subsection{Location Optimization}

Location optimization is an effective technique to further enhance the system performance.
The placement of the relay node is a practical problem in cellular deployment.
The movable relay, named as nomadic relay station, has already been proposed by the IEEE 802.16¡¯s Relay Task Group to serve a particular group servers in~\cite{WCoi}.
In these cases, the relay location optimization become an essential issue in term of system design and performance improvement.
In this subsection, we formulate the relay location optimization problem based on the the minimal SER criterion.

\textbf{\emph{Problem Formulation: }} For given total transmit power $P$, sum distance $D$, direct distance $D_{SD} < D$, power allocation ratio ${\rho _{\lambda L } } = {P_S}/P$, RSI level $\varepsilon$,  and the path loss exponent $v$ of the wireless channel, the optimal relay location ratio ${\rho _D}  = {D_{SR}}/D$  can be determined by
\begin{eqnarray}
&&\min SER({\rho _\lambda }  = {\rho _{\lambda L } },{\rho _D} ), \nonumber \\
&&\text{subject to }  1 > {\rho _D}  > 0.
\end{eqnarray}

First, we prove that the SER performance of FD relay system is convex in term of the relay location ratio ${\rho _D}$.

\textbf{\emph{Lemma 1: }}
The optimal relay location ratio ${\rho _D }$ which minimizes $SER({\rho _\lambda } ,{\rho _D} )$ is unique with predefined power allocation.

\emph{\textbf{Proof: }}
See Appendix C.$\hfill\blacksquare$

\textbf{\emph{Theorem 3: }} In the high transmit power region, the optimal location can be approximately obtained as
\begin{equation}\label{eq:dopt}
\rho _{DL} ^* = \frac{{{D_{SR}}}}{D} \approx \frac{1}{{1 + {{(\frac{{1 + \varepsilon {P_R}}}{{{P_S}}}{P_R})}^{\frac{1}{{v - 1}}}}}}.
\end{equation}

From \eqref{eq:dopt}, we can note that the optimal relay location tends to approach the source node as the relay transmit power $P_R$ increases.
This is because the RSI increases with the relay transmit power, and the received SINR at the FD relay is degraded,
and thus to maintain the quality of the received signal at the relay, the relay has to move close to the source node to enhance the channel $h_{SR}$.
Similarly, as the source transmit power reduces, the optimal relay location also moves close to the source.

In the high SNR region, the SER of the FD relay system with location optimization in \eqref{eq:dopt} can be expressed as
\begin{equation}\label{serloo}
SE{R_{LO}} \approx \frac{1}{2} - \kappa \left( {\frac{\beta }{2} + \frac{{(\frac{1}{{{\rho _{\lambda L}}P}} + \frac{{{{\overline {{\rho _\lambda }} }_L}}}{{{\rho _{\lambda L}}}}\varepsilon ){D^v}}}{{{{(1 + {{(\frac{{{{\overline {{\rho _\lambda }} }_L}}}{{{\rho _{\lambda L}}}} + \varepsilon \frac{{{{\overline {{\rho _\lambda }} }_L}^2}}{{{\rho _{\lambda L}}}}P)}^{\frac{1}{{v - 1}}}})}^{v - 1}}}}} \right),
\end{equation}
where $\kappa = \frac{{\alpha \sqrt \beta  }}{{2\sqrt {2\pi } }}\Gamma (\frac{1}{2})$, ${{\bar {\rho _\lambda } }_L} = 1 - {\rho _{\lambda L } }$.

When SNR goes infinite, $SER_{LO,{SNR -  > \infty }}=(1-\alpha)/2$, indicating that the performance floor in high SNR region can be removed with location optimization.

\subsection{Power Allocation}
Deploying power allocation in the FD relay network is effective in mitigating the impact of RSI on the system performance.
In this subsection, we design the power allocation algorithm based on the minimal SER criterion.

\textbf{\emph{Problem Formulation: }} For given total transmit power $P$, sum distance $D$, direct distance $D_{SD} < D$, relay location $\rho _{DP}= {\rho _D} = {D_{SR}}/D$, RSI level $\varepsilon$,  and the path loss exponent $v$ of the wireless channel, the optimal power allocation ratio ${\rho _\lambda } = {P_S}/P$ can be determined by
\begin{eqnarray}
&&\min SER({\rho _\lambda } ,{\rho _D}  = {\rho _{DP} }), \nonumber \\
&&\text{subject to }  1 > {\rho _\lambda } > 0.
\end{eqnarray}

First, we prove that there exists one unique power allocation ratio with respect to predefined relay location.

\textbf{\emph{Lemma 2:}}
The optimal power allocation ratio ${\rho _\lambda }$ which minimizes $SER({\rho _\lambda } ,{\rho _D} )$ is unique with predefined relay location.

\emph{\textbf{Proof: }}
See Appendix D.$\hfill\blacksquare$

\textbf{\emph{Theorem 4: }} In the high transmit power region, the optimal power allocation can be approximately obtained as
\begin{equation}\label{eq:popt}
\rho _{\lambda P}^* = \frac{{{P_S}}}{P} \approx \frac{1}{{1 + {{(\frac{{D_{RD}^v}}{{D_{SR}^v + P{B_0}\varepsilon D_{SR}^v}})}^{\frac{1}{2}}}}}.
\end{equation}

From \eqref{eq:popt}, we can observe that the source transmit power increases when the relay moves away from the source.
It can be explained that the RSI increases as the channel between the source and relay becomes worse.
The source transmit power needs to be increased in order to sustain the quality of the relay's received signal.
In addition, as RSI level $\varepsilon$ increases, the source transmit power also needs to be increased in order to improve the received SINR at the relay.

In the high SNR region, the SER of the FD relay system with power allocation in \eqref{eq:popt} can be derived as
\begin{align}\label{withpaser}
&SER{_{PA}}\approx \frac{1}{2} - \frac{{\alpha \sqrt \beta  }}{{2\sqrt {2\pi } }}\Gamma (\frac{1}{2})(\frac{\beta }{2} + \frac{1}{P}{D^v}({\rho _{DP}}^v + {(1- {\rho _{DP}})^v}  \nonumber\\
 &\hspace{15mm} + 2{\rho _{DP}}^{v/2}{(1- {\rho _{DP}})^{v/2}}\sqrt {P \varepsilon + 1} ){)^{ - \frac{1}{2}}},
\end{align}
When SNR goes to infinity, $SER_{PA,{SNR -  > \infty }}=(1-\alpha)/2$, indicating that the performance floor in high SNR region can be removed with power allocation.

\subsection{Joint Power Allocation And Relay Location Optimization}
In this subsection, we jointly optimize the relay location and power allocation to minimize the SER. The SER minimization problem is formulated under total transmit power constraint.

\textbf{\emph{Problem Formulation: }} For any total transmit power $P$, sum distance $D$, direct distance $D_{SD} < D$, RSI level $\varepsilon$  and the path loss exponent $v$ of the wireless channel, the optimal relay location ratio ${\rho _D}  = {D_{SR}}/D$  and power allocation ratio ${\rho _\lambda }  = {P_S}/P$ can be determined by
\begin{eqnarray}
&&\min SER({\rho _\lambda } ,{\rho _D} ), \nonumber \\
&&\text{subject to } 1 > {\rho _\lambda }  > 0, 1 > {\rho _D}  > 0.
\end{eqnarray}

The Hessian matrix of $SER({\rho _\lambda } ,{\rho _D} )$ is not positive definite under constraints $1 > {\rho _\lambda }  > 0, 1 > {\rho _D}  > 0$.
Therefore,  the SER function is not a convex function of variables ${\rho _\lambda }, {\rho _D}$ and the global minimizer for the joint optimization problem may not be unique.
With the following proposition, all the global minimizers for the joint optimization problem can be provided.

\textbf{Proposition 1: }
For any total transmit power $P$, RSI level $\varepsilon$  and the path loss exponent $v$ of the wireless channel, all the global minimizers $({\rho _\lambda} , {\rho _D } )$ for the joint optimization problem can be obtained by solving the following equations 
\begin{eqnarray}\label{eq:pro1}
&&{(1 + \varepsilon P\bar {{\rho _\lambda }} )^v}{(\frac{1}{{\bar {{\rho _\lambda }} }} - 1)^{v - 2}} = {(1 + P\varepsilon )^{v - 1}}, 0< \bar {{\rho _\lambda }} < 1\nonumber\\
 &&{\rho _D} = 1/(1 + {(\frac{{1 + \varepsilon \bar {{\rho _\lambda }} P}}{{{1 - \bar{\rho _\lambda }}}}\bar {{\rho _\lambda }} )^{\frac{1}{{v - 1}}}}),
\end{eqnarray}
where $\bar {{\rho _\lambda }} = 1- {\rho _\lambda }$.

\emph{\textbf{Proof: }}
See Appendix E.$\hfill\blacksquare$

The first equation in Proposition 1 is a $2v - 2$ order equation.
When $v$ increases, the number of solutions to the equation increases.
While the exact solutions for \eqref{eq:pro1} can be obtained via mathematical tools, the solutions can be complicated enough to provide any insight when $v$ is high.

In wireless communications, path loss is represented by the path loss exponent, whose value is normally in the range of $2$ to $4$.
In this paper, we use $v = 3$ for example, which is typical in wireless communication network, e.g., indoor propagation, office with partition.

When $v=3$, the solutions for \eqref{eq:pro1} are given by
\begin{align}\label{eq:jsolute}
  \rho _{\lambda {J_1}}^* &= \frac{{\sqrt {1{\rm{ + }}\varepsilon P} }}{{\sqrt {1{\rm{ + }}\varepsilon P}  + 1}},\nonumber\\
  \rho _{\lambda {J_{2,3}}}^* &= \frac{{1 + \varepsilon P \pm \sqrt { - 3 - 2\varepsilon P + {\varepsilon ^2}{P^2}} }}{{2\varepsilon P}}, {\rm{(}}\varepsilon P \ge 3{\rm{)}}{\rm{.}}
\end{align}
With \eqref{eq:j2} and \eqref{eq:jsolute}, the corresponding optimized location $\rho _{DJ}^*$ can be derived. The minimized SER is given by
\begin{equation}\label{eq:serminn}
  SE{R_{Joint}} = \min \{ SER(\rho _{\lambda {J_i}}^*,\rho _{D{J_i}}^*),i = 1,2,3\},
\end{equation}

When $v = 2$, for the joint optimization,  the solution to \eqref{eq:pro1} can be derived as $\rho _{\lambda J}^ *  = \frac{{\sqrt {1 + \varepsilon P} }}{{\sqrt {1 + \varepsilon P}  + 1}},\rho _{DJ}^ *  = \frac{1}{2}$.

For the sequential optimization, by substituting $\rho _{DL} ^*$  in \eqref{eq:dopt} into \eqref{eq:f}, the value of $f({\rho _\lambda } ,{\rho _D} )$ can be further optimized. When $v=2$, the corresponding optimal power allocation ratio $\rho _{\lambda L } ^*$ can be derived as $\rho _{\lambda L } ^* = \frac{{\sqrt {1 + \varepsilon P} }}{{\sqrt {1 + \varepsilon P}  + 1}}$, indicating that the SER performance can be further optimized if the predefined power allocation ratio is fixed at $\rho _{\lambda L } ^* = \frac{{\sqrt {1 + \varepsilon P} }}{{\sqrt {1 + \varepsilon P}  + 1}}$.

Similarly, by substituting $\rho _{\lambda P}^*$ in \eqref{eq:popt} into \eqref{eq:f}, the optimized relay location can be derived as $\rho _{DP}^* = \frac{1}{2}$ when $v = 2$, indicating that the optimal relay location is equidistant from the source and destination, and not affected by the total transmit power and RSI.

Therefore, the optimized relay location and power allocation ratio obtained from the joint optimization and sequential optimizations are the same when $v = 2$.

\textbf{Proposition 2: }
For any total transmit power $P$, RSI level $\varepsilon$  and the path loss exponent $v$ of the wireless channel, a particular solution to the joint optimization problem in Proposition 1 can be calculated as
\begin{equation}\label{eq:parsolution}
  \rho _{\lambda J}^ *  = \frac{{\sqrt {1 + \varepsilon P} }}{{\sqrt {1 + \varepsilon P}  + 1}},\rho _{DJ}^ *  = \frac{1}{2}.
\end{equation}

\emph{\textbf{Proof: }}
The solution can be readily verified.
Therefore, the joint optimization problem in Proposition 1 has at least one feasible solution.$\hfill\blacksquare$

The solution in \eqref{eq:parsolution} is a suboptimal solution to the joint optimization problem.
From \eqref{eq:parsolution}, the source transmit power increases as RSI level $\varepsilon$ increases and the optimized relay location is equidistant from the source and destination.
This is because the received SINR at the FD relay is degraded with the increase of RSI, and the source transmit power needs to be increased to maintain the quality of the relay's received signal.

\section{Simulation Results}

In this section, we present the numerical results for the FD AF relaying systems.
Without loss of generality, we assume the pathloss exponent $v=3$, and a BPSK modulation is used in the outage probability and SER performance evaluations.

\begin{figure}[!t]
\centering
\includegraphics[width=5in]{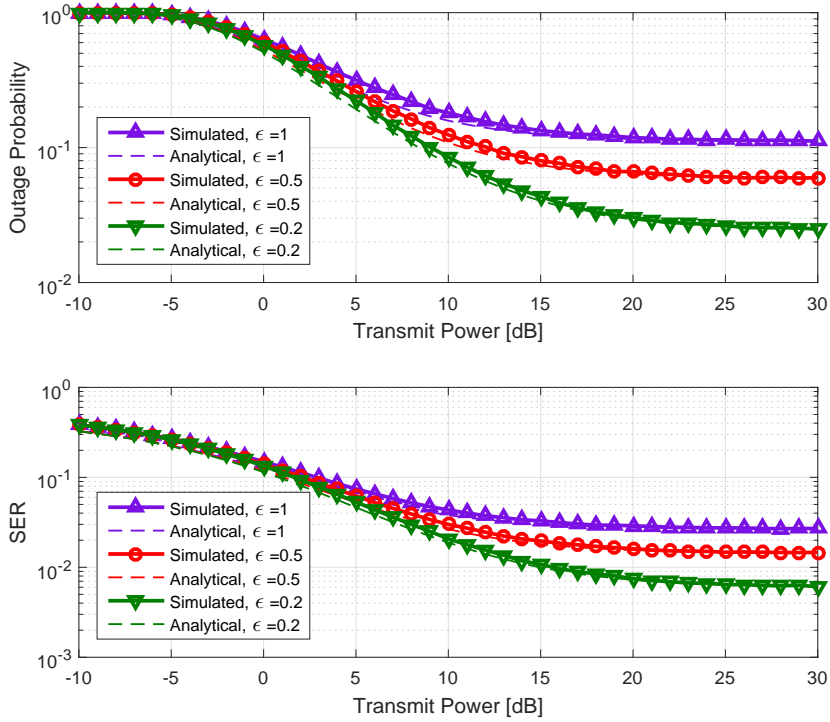}
\vspace{-1mm}
\caption{\label{fig:perfor}The outage probability and SER performance for different RSI levels, $(P_S /P = 0.5, D_{SR}/D = 0.5, v = 3)$.}
\vspace{-4mm}
\end{figure}
In Fig.~\ref{fig:perfor}, we plot the numerical and analytical results of the outage probability and SER performance for the relay system with one FD relay located equidistant from the source and destination.
The simulated outage probability and SER curves tightly match with the expressions in~\eqref{eq:CDF_onerelay} and~\eqref{eq:ser_approx}.
From the figure, we observe that both the outage probability and SER performance improve with the decrease of RSI.
The performance floor of the SER curve coincides with the result computed by the equation~\eqref{eq:performance floor}.

\begin{figure}[!t]
\centering
\includegraphics[width=5in]{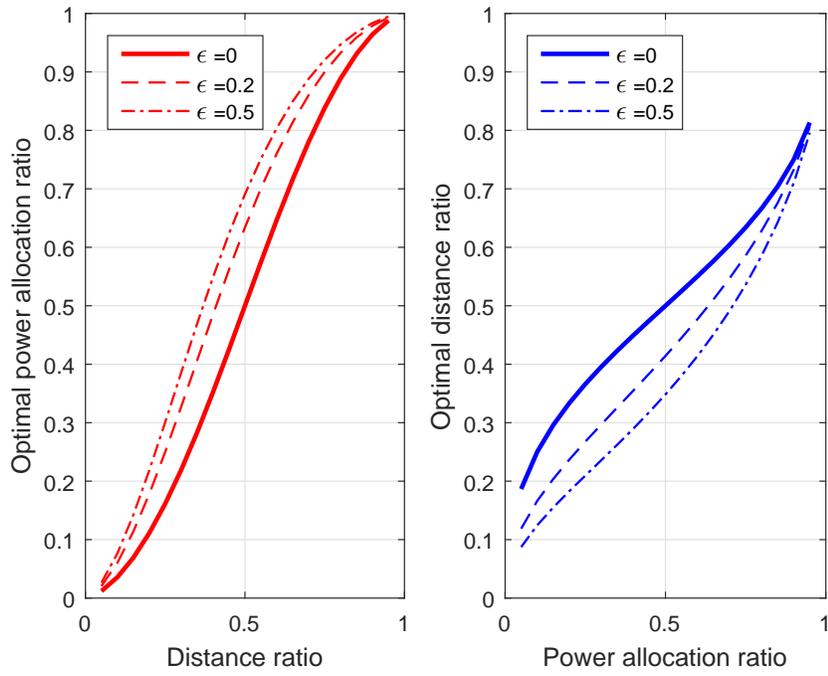}
\vspace{-1mm}
\caption{\label{fig:optimalfor}The optimal power/distance ratio for distance/power ratio, $(P = 10dB, v = 3)$. }
\vspace{-4mm}
\end{figure}

Fig.~\ref{fig:optimalfor} depicts the optimal power/location ratio for the relay system with different location/power ratios.
From the figure, we observe that more transmit power is allocated to the source node as the relay moves away from the source.
For the system with fixed relay location, as RSI increases, the optimal transmit power allocated to the source also increases.
For the system with fixed power allocation, the optimal relay moves close to the source as RSI increases.
This is because the received power at the relay needs to be increased in order to maintain the received SINR at relay due to the increase of RSI.

\begin{figure}[!t]
\centering
\includegraphics[width=5in]{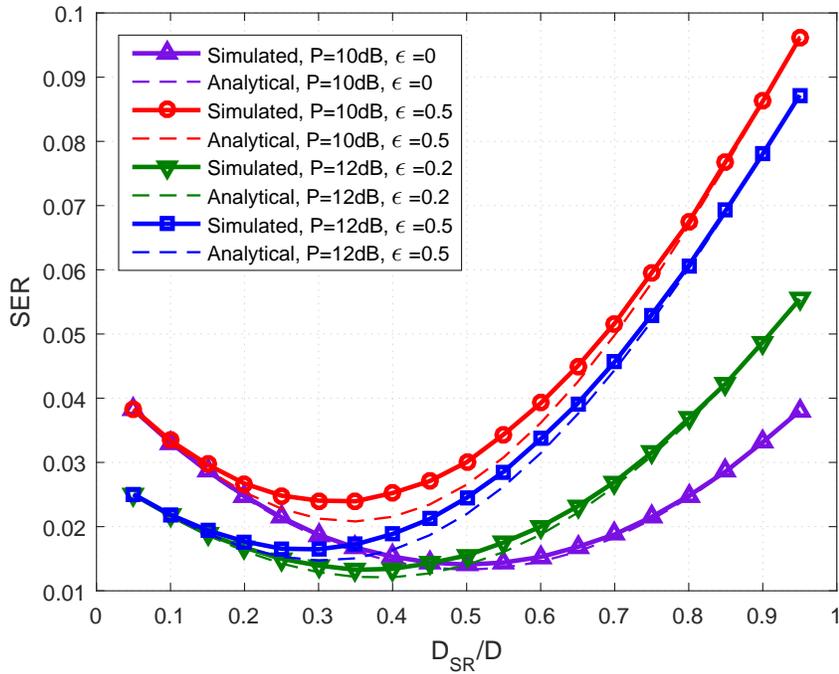}
\vspace{-1mm}
\caption{\label{fig:loc}The average SER comparisons with different relay locations, $(P_S /P = 0.5, v = 3)$.}
\vspace{-4mm}
\end{figure}

Fig.~\ref{fig:loc} investigates the impact of the relay location on SER performance when the transmit power at the source and relay is set as equal.
The optimal relay location derived by Theorem 3 is a suboptimal solution for the location optimization problem, and approaches the optimal one in the high transmit power.
From this figure, we observe that when RSI increases, the optimized location of FD relay moves close to the source.
This is because that the received signal at the relay gets worse with RSI increases, the relay has to move close to the source to maintain received signal quality. 
When the interference cancellation is perfect at the relay, the optimized relay location is equidistant from the source and the destination.
This is consistent with our analysis in Theorem 3.

\begin{figure}[!t]
\centering
\includegraphics[width=5in]{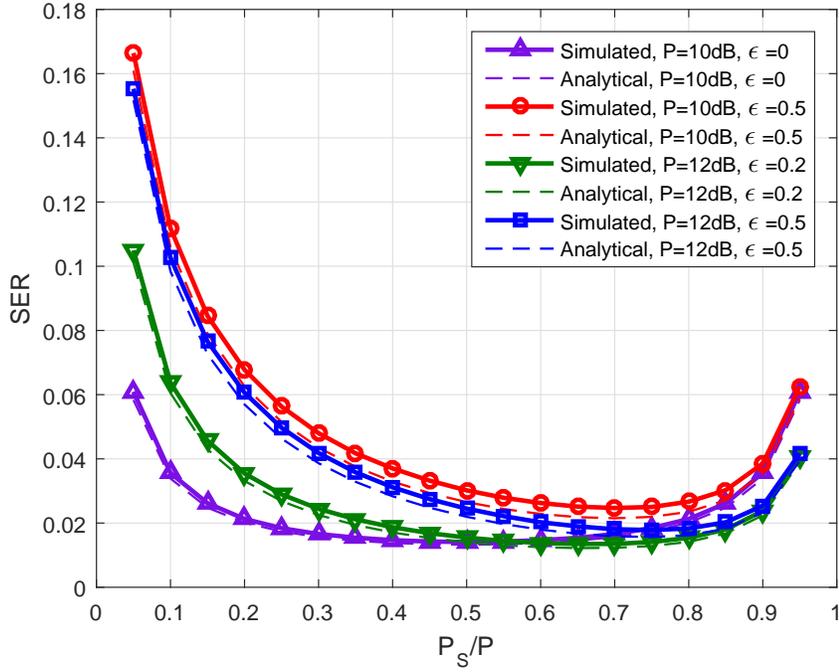}
\vspace{-1mm}
\caption{\label{fig:pa}The average SER comparisons with different power allocations, $(D_{SR}/D = 0.5, v = 3)$.}
\vspace{-4mm}
\end{figure}

Fig.~\ref{fig:pa} shows the SER performance with different power allocations.
The relay location is equidistant from the source and destination.
The optimal power allocation in Theorem 4 is a suboptimal solution and approaches the optimal one in the high transmit power.
We can observe that the source transmit power increases as RSI increases.
Due to the increase of RSI, the received signal at the relay gets worse.
In order to improve the SINR of received signal at the relay, the source transmit power needs to be increased.
As RSI decreases, the optimal power allocation ratio ${\rho _\lambda }$ approaches $\frac{1}{2}$.
If the self interference cancellation is ideal, the total transmit power is evenly divided between the source and relay, which is consistent with our analysis in Theorem 4.

\begin{figure}[!t]
\centering
\includegraphics[width=5in]{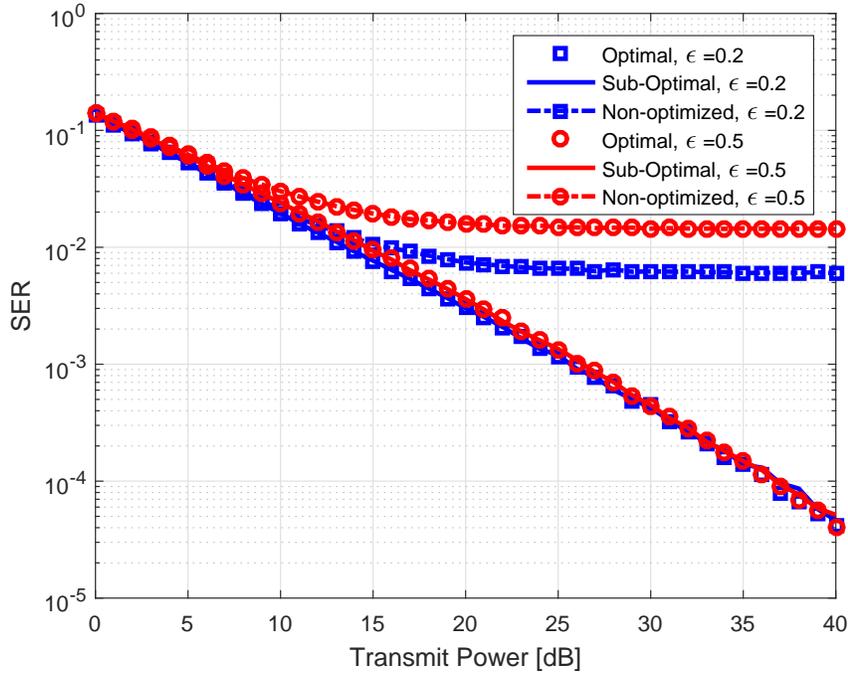}
\vspace{-1mm}
\caption{\label{fig:ser_lo_cmp}The average SER performance with and without the relay location optimization, $(P_S /P = 0.5, v = 3)$.}
\vspace{-4mm}
\end{figure}

Fig.~\ref{fig:ser_lo_cmp} compares the SER performance with the optimal and the derived suboptimal relay location.
The performance of the non-optimized relay system with relay located equidistant from the source and destination, is also illustrated for comparison.
From this figure, we can see that the derived suboptimal solution approaches the optimal one in the whole SNR regime and outperforms the non-optimized relay location.
The SER performance gain increases as the SNR increases.
In the high SNR region, the non-optimized relay scheme encounters the performance floor, which is caused by RSI.
However, the performance floor can be significantly reduced with both optimal and suboptimal relay location optimization.

\begin{figure}[!t]
\centering
\includegraphics[width=5in]{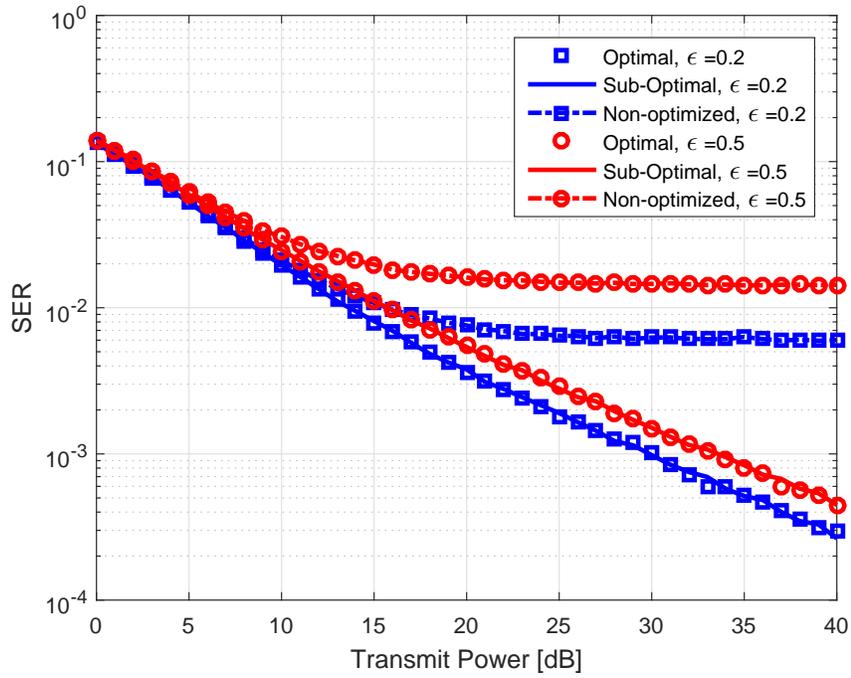}
\vspace{-1mm}
\caption{\label{fig:ser_pa_cmp}The average SER performance with and without power allocation, $(D_{SR}/D = 0.5, v = 3)$.}
\vspace{-4mm}
\end{figure}

Fig.~\ref{fig:ser_pa_cmp} compares the SER performance with the optimal and suboptimal power allocation.
The SER performance of the non-optimized system, where the transmit power is evenly divided between the source and relay, is also plotted for comparison.
From the figure, we observe that the gap between the optimal and our proposed scheme is very small, even in the low SNR region.
The SER performance of the FD relay system without power allocation is worse than the optimized system, and the performance gap is very large in the high SNR region due to the increasing RSI.
In the high SNR region, the performance floor is removed by deploying either optimal or suboptimal power allocation scheme.

\begin{figure}[!t]
\centering
\includegraphics[width=5in]{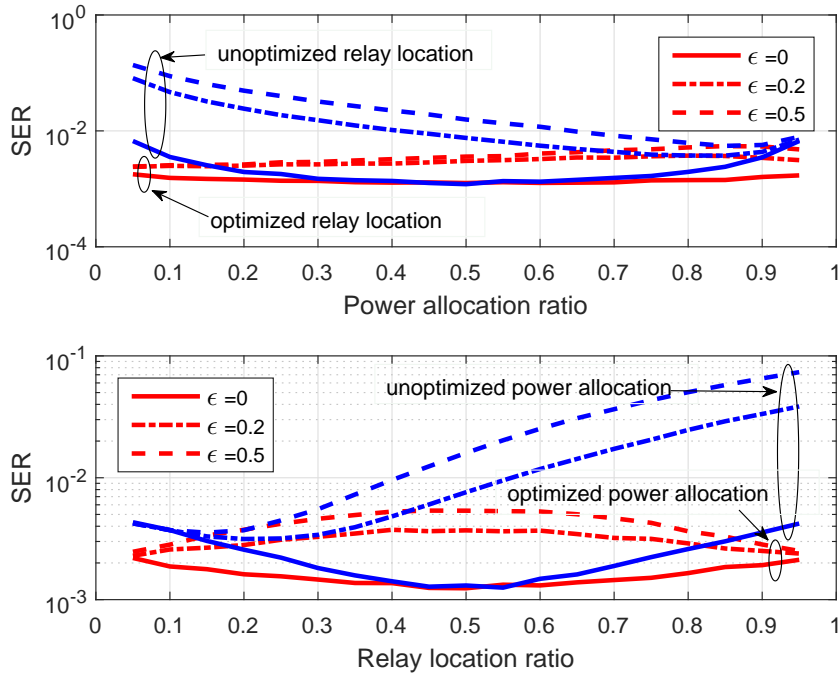}
\vspace{-1mm}
\caption{\label{fig:ser_lopa_optimized}The average SER performance with and without power/location optimization,  $(P = 20dB, v = 3)$.}
\vspace{-4mm}
\end{figure}
Fig.~\ref{fig:ser_lopa_optimized} compares the simulated SER performance between the FD relay systems with and without location/power optimization.
It is shown that the FD relay system with optimized relay location outperforms the un-optimized system with the same power allocation and RSI.
We can also observe that the SER performance curve is more flat for the system with optimized locations, indicating that the channel links between the source/relay and relay/destination become more balanced than the un-optimized links.
Similar results can be observed from the SER performance comparison between the FD relay systems with and without power allocation.

\begin{figure}[!t]
\centering
\includegraphics[width=5in]{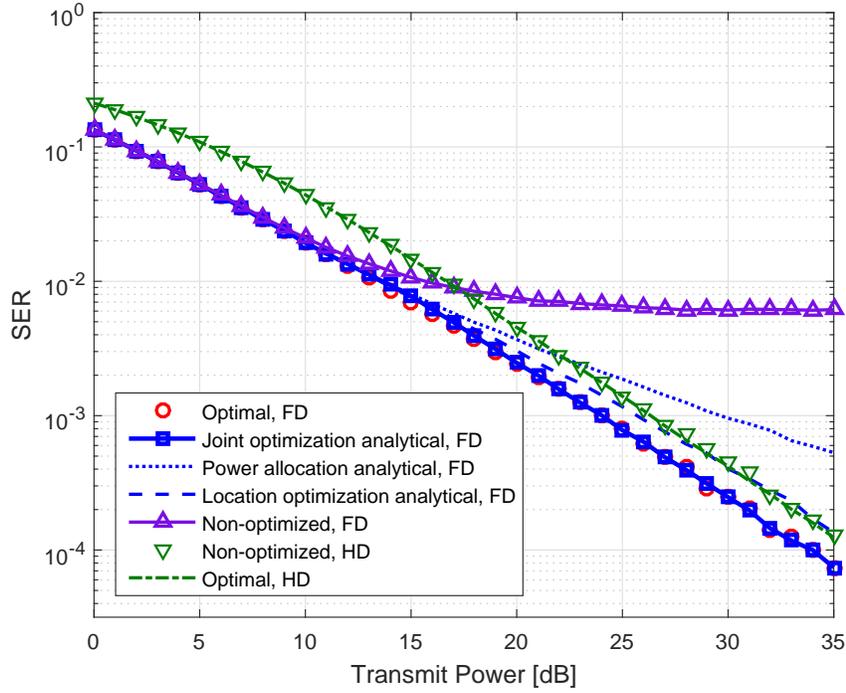}
\vspace{-1mm}
\caption{\label{fig:ser_lo_pa_cmp}The average SER comparisons between different optimization schemes, $(\varepsilon =0.2, v = 3)$.}
\vspace{-4mm}
\end{figure}

Fig.~\ref{fig:ser_lo_pa_cmp} compares the simulated SER performance of schemes with joint optimization, power allocation only, location optimization only, respectively.
For the power allocation optimization only scheme, the relay is equidistant from the source and destination.
For the relay location optimization only scheme, the transmit power at the source and relay are set as equal.
The SER performance of non-optimized FD and HD system is also plotted for comparison where $P_S/P = 1/2, {D_{SR}} / D = 1/2$.
The SER performance of suboptimal solutions obtained in \eqref{eq:jsolute} approaches that of the optimal solution.
From the figure, we can observe that the joint optimization can bring considerable gains compared to the location optimization only and power allocation only schemes, especially in high SNR region.
Therefore, joint optimization is essential to obtain the minimum SER performance.
It is also shown that all the three schemes remove the performance floor caused by imperfect self interference cancellation.

\begin{figure}[!t]
\centering
\includegraphics[width=5in]{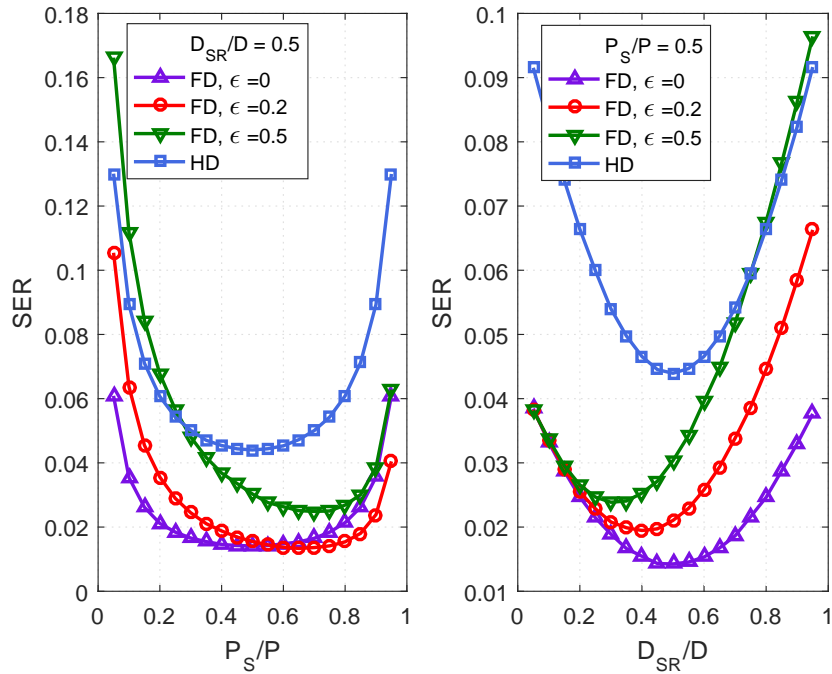}
\vspace{-1mm}
\caption{\label{fig:ser_fdhd}The average SER comparisons between FD and HD with different power allocations or relay locations, $(P = 10dB, v = 3)$.}
\vspace{-4mm}
\end{figure}

Fig.~\ref{fig:ser_fdhd} compares the simulated SER performance of FD system and HD system with different power allocations or relay locations at different RSI levels.
It is shown that the optimal power allocation/relay location and the corresponding optimal SER performance of FD mode changes with the RSI.
As RSI increases, the optimized source transmit power increases, the optimized relay location moves close to the source and the corresponding optimal SER performance decreases.
However, the optimal power allocation of the HD mode is fixed once the relay location is set and is not influenced by the RSI, and vise versa.
Therefore, the RSI causes additional complexity in the power allocation and location optimization in the FD system.
From Fig.~\ref{fig:ser_lo_pa_cmp}, we can also observe that either optimized power allocation or optimized relay location can remove the performance floor at high SNR and considerably increases the SER performance in FD system.
While in the HD system, the optimal power allocation and location optimization provides slightly better SER performance compared with non-optimized performance.
Therefore, power allocation and location optimization are more important to obtain better system performance in FD system than in HD system due to the RSI.

\subsection{Comparison of three schemes}

In this paper, we analyze three optimization schemes including pure power allocation, pure location optimization and joint power and location optimization.
Both three schemes significantly reduce the performance floor caused by RSI and considerably improves the system performance compared with the non-optimized full-duplex relay system.
It is shown in Fig.~\ref{fig:ser_lo_pa_cmp} that the joint optimization is essential to obtain the minimum SER performance.
It also needs to be noted that pure optimizations are useful in practical scenarios with different restrictions, requirements and deployment issues.
The joint optimization is under the assumption that the power and location can be jointly adjusted based on the CSI.
In practical relay transmission situations where either power or location is predefined or restricted, only pure location or pure power optimization can be adopted to improve the performance.
Also, the solutions of pure power or pure location optimization \eqref{eq:dopt}, \eqref{eq:popt} can be readily applied in practical scenarios, while additional calculations may be required for deployment in joint optimization issues.
In addition, there is not a considerable increase from the SER performance perspective of joint optimization compared with pure location and power optimization, which is observed from Fig.~\ref{fig:ser_lo_pa_cmp}.
In some situations where the minimum SER or perfect performance are not essentially required, pure power and location optimization can be adopted.

\section{Conclusions}
In this paper, power allocation and relay location optimization for the FD AF relay system were investigated.
The asymptotic expressions of the outage probability and SER performance were derived.
With the asymptotic SER, the relay location optimization, power allocation and joint optimization problem were formulated based on the minimal SER criterion.
The derived suboptimal solutions perform very close to the optimal ones in terms of SER.
The results showed that RSI has a large influence on the optimized results.
Both the relay location optimization and power allocation provides considerable SER gain.
Moreover, the performance floor caused by RSI due to imperfect self interference cancellation can be significantly reduced with either power allocation or location optimization.
It was also shown that joint location and power optimization can provide additional gains compared to the pure power allocation or location optimization.

\appendices
\section{Proof of Theorem 1}

In the high transmit power, the end-to-end SINR in~\eqref{eq:e2eSINR} can be rewritten as
\begin{equation}
\label{eq:sinr_1}
{\gamma _{SINR}} = \frac{\frac{\gamma_{SR}}{\gamma_{LI}+1}\gamma_{RD}} {\frac{\gamma_{SR}}{\gamma_{LI}+1}+\gamma_{RD}+1}
\approx \frac{X \gamma_{RD}}{X+\gamma_{RD}},
\end{equation}
where $X = \frac{\gamma_{SR}}{\gamma_{LI}+1}$.
The distribution of $X$ is given by $F_{X} (x) = 1 - \frac{1}{1 + \eta x} e^{-\frac{1}{\lambda_{SR}}x}$, where $\eta = \lambda_{LI}/\lambda_{SR}$.

Note that, due to the special distribution of the variable $X$, the performance analysis for the HD relay systems cannot be directly applied in the FD relay systems.
Therefore, in this paper, we derive an asymptotic outage probability of the FD relay systems from another point of view.
The CDF of the end-to-end SINR~\eqref{eq:sinr_1} is expressed as
\begin{eqnarray}
\label{eq:CDF_integral}
\Pr({\gamma _{SINR}} > x) = \Pr \Big\{  (X-x)(\gamma_{RD}-x) > x^2 \Big\} \nonumber \\
= \frac{1}{\lambda_{RD}} \int\limits_x^\infty \frac{e^{-\frac{1}{\lambda_{SR}}(x+\frac{x^2}{\gamma_{RD}-x})-\frac{1}{\lambda_{RD}}\gamma_{RD}}} {1+\eta (x+\frac{x^2}{\gamma_{RD}-x})} d\gamma_{RD}.
\end{eqnarray}

To the best of the authors' knowledge, the integral of the exact CDF distribution does not have a closed-form solution.

The integral~\eqref{eq:CDF_integral} can be separated into two parts.
\begin{align}
\Pr({\gamma _{SINR}}>x) &= \frac{C}{{{\lambda _{RD}}}}\int\limits_0^\infty  {\frac{{{e^{ - \frac{{{x^2}}}{{{\lambda _{SR}}t}} - \frac{t}{{{\lambda _{RD}}}}}}}}{{1 + \eta x}}} dt{\rm{ }} \nonumber\\
&\hspace{3mm}- \frac{C}{{{\lambda _{RD}}}}\int\limits_0^\infty  {\frac{{{x^2}\eta {e^{ - \frac{{{x^2}}}{{{\lambda _{SR}}t}} - \frac{t}{{{\lambda _{RD}}}}}}}}{{(1 + \eta x)(t + \eta xt + {x^2}\eta )}}} dt \nonumber \\
&= I_1 - I_2,
\end{align}
where $C = e^{ - (\frac{1}{\lambda_{SR}} + \frac{1}{\lambda_{RD}}) x}$.
The first part $I_1$ of this equation can be calculated
\begin{equation}
I_1 = \frac{C}{1+\eta x} \frac{2x}{\sqrt{\lambda_{SR}\lambda_{RD}}} K_1 \left(\frac{2x}{\sqrt{\lambda_{SR}\lambda_{RD}}}\right).
\end{equation}
According to the definition of function $K_1(\cdot)$, $x K_1(x) \to 1$ when $x \to 0$~\cite{Abramowitz}.
Therefore, in the high transmit power conditions, $I_1 \to C/(1+\eta x)$.

The second part $I_2$ is upper bounded by
\begin{eqnarray}
I_2 &<& \frac{C}{(1+\eta x)\lambda_{RD}} \int\limits_0^{\infty} \frac{1}{1+\frac{1+\eta x}{ \eta x^2 }t } e^{-\frac{t}{\lambda_{RD}}} d t \nonumber\\
&=& \frac{C\eta x^2}{\lambda_{RD}(1+\eta x)} e^{\frac{\eta x^2 }{\lambda_{RD}(1+\eta x)}} E_1 \left[\frac{1}{\lambda_{RD}} \frac{ \eta x^2}{1+\eta x} \right],
\end{eqnarray}
where the function $E_1(\cdot)$ is the exponential integral~\cite{Abramowitz}.
According to the properties of exponential integral, $xE_1(x)\to 0$ when $x \to 0$~\cite{Gradshteyn94}.
When the transmit power increases or RSI decreases, the integral $I_2$ approaches zero.
Further, $I_2$ can be dropped under high transmit power or low self interference conditions, and thus integral \eqref{eq:CDF_integral} is simplified into a closed-formed expression.

Therefore, the asymptotic expression for the outage probability can be obtained as
\begin{align}\label{eq:CDF_asym}
&F(x) = 1-\Pr({\gamma _{SINR}}>x) \nonumber\\
&\approx 1 - \frac{{{e^{ - (\frac{1}{{{\lambda _{SR}}}} + \frac{1}{{{\lambda _{RD}}}})x}}}}{{1 + \eta x}}\frac{{2x}}{{\sqrt {{\lambda _{SR}}{\lambda _{RD}}} }}{K_1}\left( {\frac{{2x}}{{\sqrt {{\lambda _{SR}}{\lambda _{RD}}} }}} \right).
\end{align}

The major difference of our paper and \cite{Antenna}  lies in different methods calculating CDF.
In our paper, we use $F(x) \approx 1- I_1$, while the researchers used $P_{MM} \ge \Pr \left\{ {\min (\frac{{\gamma _{SR}^{I,J}}}{{\gamma _{RR}^{I,K} + 1}},\gamma _{RD}^{K,L}) < {\gamma _T}} \right\}$ in \cite[eq.(50)]{Antenna}. The two methods are essentially different and the derived asymptotic expression \eqref{eq:CDF_asym} in our paper is much closer to the exact outage probability, especially in the low to moderate SNR regions. 

\section{Proof of Theorem~2}

To compute the integral, we need to introduce an approximation for $1/(1+x)$, and its Taylor expansion around $x = 0$ is given by $1/(1+x) = 1-x+x^2-x^3+ \dots $ when $|x|<1$.
In this paper, we introduce an approximation for $1/(1+x)$ based on the Taylor expansion to compute the average SER.
\begin{equation}\label{eq:approx_frac}
\frac{1}{1+x} \approx \sum_{i = 0}^{\infty} A_i x^{2i} \exp\left(-B_i x\right).
\end{equation}
The parameters $A_i$ and $B_i$ are fixed to guarantee that the factors of the $(2i+1)$th and $(2i+2)$th items, $x^{2i+1}$ and $x^{2i+2}$, in ~\eqref{eq:approx_frac} are equal to those in the Taylor expression.
Therefore, the parameters $A_i$ and $B_i$ can be computed sequentially based on the previous terms,
\begin{equation}
\label{eq:para_ser}
A_i = 1 - \sum_{j = 0}^{i-1} \frac{A_j B_j^{2i-2j}}{(2i-2j)!}, B_i = \Big(1 - \sum_{j = 0}^{i-1} \frac{A_j B_j^{2i-2j+1}}{(2i-2j+1)!}\Big)/A_i.
\end{equation}
When x is close to zero, the approximation is accurate.
When x is high, the effect of the estimation error is very small due to the quickly decreasing property of Q-Function.
In this paper, we use the first three $({A_i},{B_i})$ pairs for approximation, which are $(1, 1)$, $(1/2, 5/3)$, $(19/72, 1963/855)$.

Then with the help of the formula~\cite[eq.(6.621.3)]{Gradshteyn94},
\begin{align}
&\int \limits_0^\infty x^{\mu-1} e^{-\alpha x} K_{\nu}(\beta x) d x
= \frac{\sqrt{\pi}(2\beta)^\nu}{(\alpha+\beta)^{\mu+\nu}} \nonumber \\
&\times\hspace{-1mm}  \frac{\Gamma(\mu+\nu)\Gamma(\mu-\nu)}{\Gamma(\mu+\frac{1}{2})}
F(\mu+\nu, \nu+\frac{1}{2}; \mu+\frac{1}{2}; \frac{\alpha-\beta}{\alpha+\beta}),
\end{align}
the theorem can be proved.

\section{Proof of Lemma~1}
Based on the approximate expression of SER performance of FD relay system in the high transmit power region, we can prove the optimal distance ratio ${\rho _D }$ which minimizes $SER({\rho _\lambda } ,{\rho _D} )$ is unique with predefined power allocation.

The second derivative of function $f({\rho _\lambda },{\rho _D}) = \frac{\beta }{2} + \frac{{1 + \varepsilon {P_R}}}{{{P_S}}}{D^v}{\rho _D}^v + \frac{1}{{{P_R}}}{D^v}{\left( {1 - {\rho _D}} \right)^v}$ with respect to $\rho_D$ can be derived as
\begin{equation}
\frac{{{\partial ^2}f({\rho _\lambda },{\rho _D})}}{{{\partial ^2}{\rho _D}}}\hspace{-1mm} = {D^v}( {{v^2} -\hspace{-1mm} 1})( {\frac{{1 + \varepsilon {P_R}}}{{{P_S}}}{\rho _D}^{v - 2} +\hspace{-1mm} \frac{{{{\left( {1 - {\rho _D}} \right)}^{v - 2}}}}{{{P_R}}}} ),
\end{equation}
With $1 > {\rho _D }  > 0$ and the path loss exponent $v > 1$,  the second derivative of $f({\rho _\lambda } ,{\rho _D} )$ with respect to ${\rho _D }$ is positive.
Therefore, $f({\rho _\lambda } ,{\rho _D} )$ is convex in term of the distance ratio ${\rho _D }$.

In the high transmit power region, the SER performance of the FD relay system can be approximated as
\begin{equation}\label{eq:serhigh}
 SER \approx \frac{1}{2} - \frac{{\alpha \sqrt \beta  }}{{2\sqrt {2\pi } }}\Gamma (\frac{1}{2}){(f({\rho _\lambda } ,{\rho _D} ))^{ - \frac{1}{2}}},
\end{equation}
As shown in \eqref{eq:serhigh}, SER is a monotonically increasing function with respect to $f({\rho _\lambda } ,{\rho _D} )$. As function $f({\rho _\lambda } ,{\rho _D} )$ is a convex function in term of the distance ratio ${\rho _D }$, there exists only one optimal distance ratio ${\rho _D }$ which satisfies ${\rho _D}  = \arg \min SER({\rho _\lambda } ,{\rho _D} )$.

\section{Proof of Lemma~2}
The second derivative of function $f({\rho _\lambda } ,{\rho _D} )$ with respect to ${\rho _\lambda }$ can be derived as
\begin{equation}\label{eq:fderiv}
  \frac{{{\partial ^2}f({\rho _\lambda } ,{\rho _D} )}}{{{\partial ^2}{\rho _\lambda } }} = (\varepsilon D_{SR}^v + \frac{{D_{SR}^v}}{P})\frac{2}{{{{\rho _\lambda } ^3}}} + (\frac{{D_{RD}^v}}{P})\frac{2}{{{{(1 - {\rho _\lambda } )}^3}}}.
\end{equation}
With $1 > {\rho _\lambda }  > 0$,  the second derivative of $f({\rho _\lambda } ,{\rho _D} )$ with respect to ${\rho _\lambda }$ is positive. Therefore, $f({\rho _\lambda } ,{\rho _D} )$ is convex in term of the power allocation ratio ${\rho _\lambda }$.

In the high transmit power region, the SER performance of the FD relay system can be approximated as
\eqref{eq:serhigh}, SER is a monotonically increasing function with respect to $f({\rho _\lambda } ,{\rho _D} )$. As function $f({\rho _\lambda } ,{\rho _D} )$ is a convex function in term of the power allocation ratio ${\rho _\lambda }$, there exists only one optimal power allocation ratio ${\rho _\lambda }$ which satisfies ${\rho _\lambda }  = \arg \min SER({\rho _\lambda } ,{\rho _D} )$.

\section{Proof of Proposition~1}
With Lemma 1 and Lemma 2, all the feasible global minimizers of $SER({\rho _\lambda } ,{\rho _D} )$ can be derived by determining the $\{ \rho _{DJ}^ * ,\rho _{\lambda J}^ * \}$ which satisfies
\begin{equation}\label{eq:jointoptimization}
  \frac{{\partial f({\rho _\lambda } ,{\rho _D} )}}{{\partial {\rho _\lambda } }} = 0,\frac{{\partial f({\rho _\lambda } ,{\rho _D} )}}{{\partial {\rho _D} }} = 0, 0 < {\rho _\lambda },{\rho _D} < 1.
\end{equation}

After a few mathematical manipulations, \eqref{eq:jointoptimization} can be simplified as
\begin{equation}\label{eq:j1}
 {(1 + \varepsilon {P_R})^v}{(\frac{P}{{{P_R}}} - 1)^{v - 2}} =\hspace{-1mm} {(1 + P \varepsilon)^{v - 1}}, 0 < {P_R} < P,
\end{equation}
\begin{equation}\label{eq:j2}
  (1 + {(\frac{{1 + \varepsilon {P_R}}}{{{P_S}}}{P_R})^{\frac{1}{{v - 1}}}}){D_{SR}} = D, 0 < {D_{SR}} < D.
\end{equation}

Therefore, Proposition 1 is proved.

\end{document}